\documentclass[aps,prb,twocolumn,groupedaddress,showpacs]{revtex4}
\usepackage{graphicx}

\def\be{\begin{equation}}
\def\ee{\end{equation}}
\def\ba{\begin{eqnarray}}
\def\ea{\end{eqnarray}}

\def\bc{\begin{center}}
\def\ec{\end{center}}
\begin{document}

\title{Microwave-induced magnetotransport phenomena in two-dimensional electron systems: Importance of electrodynamic effects 
}

\author{S. A. Mikhailov}
\email[Electronic address: ]{sergey.mikhailov@mh.se}

\affiliation{Mid-Sweden University, ITM, Electronics Design Division, 851 70 Sundsvall, Sweden}

\date{\today}

\begin{abstract}
We discuss possible origins of recently discovered microwave induced photoresistance oscillations in very-high-electron-mobility two-dimensional electron systems. We show that electrodynamic effects -- the radiative decay, plasma oscillations, and retardation effects, -- are important under the experimental conditions, and that their inclusion in the theory is essential for understanding the discussed and related microwave induced magnetotransport phenomena.
\end{abstract}

\pacs{73.21.-b, 78.67.-n, 78.70.Gq, 73.43.-f}





\maketitle

\section{Introduction}

Recently discovered effects of microwave induced giant photoresistance oscillations \cite{Zudov01,Ye01} and zero-resistance states \cite{Mani02,Zudov03} in very-high-electron-mobility two-dimensional electron systems (2DES) attracted much experimental \cite{Dorozhkin03,Yang03,Mani04,Studenikin04,Mani04a,Du04,Willett03,Mani03,Kovalev03,Studenikin04b} and theoretical \cite{Andreev03,Shi03,Durst03,Bergeret03,Koulakov03,Phillips03b,Dmitriev03,Lei03a,Ryzhii03a,Ryzhii03b,Ryzhii03e,Vavilov03,Ryzhii04,Mikhailov03c,Lee03,Ryzhii03c,Klesse03,Volkov03b,Ryzhii03d,Lei03b,Dmitriev03b,Patrakov04,Dmitriev04} interest. In spite of a large number of theoretical scenarios published so far \cite{Dorozhkin03,Andreev03,Shi03,Durst03,Bergeret03,Koulakov03,Phillips03b,Dmitriev03,Lei03a,Ryzhii03a,Ryzhii03b,Ryzhii03e,Vavilov03,Ryzhii04,Mikhailov03c,Lee03,Ryzhii03c,Klesse03,Volkov03b,Ryzhii03d,Lei03b,Dmitriev03b,Patrakov04,Dmitriev04} (see also the pioneering work by Ryzhii \cite{Ryzhii70,Ryzhii86}) full understanding of the phenomenon has not yet been achieved. It is not clear, for instance, why the giant photoresistance oscillations and zero-resistance states are observed only in samples with the electron mobility exceeding $\simeq 10^7$ cm$^2$/Vs: in samples with the one-order-of-magnitude lower mobility a completely different and easily understandable behaviour was observed \cite{Vasiliadou93}. Another unclear issue is the influence of finite dimensions of the sample and plasma oscillations in it.

So far published theoretical scenarios mainly discuss the phenomenon in terms of the influence of microwaves on the probability of electron scattering or on the steady-state electron distribution function. The goal of this paper is to point to the importance of {\em electrodynamic} effects which have been ignored in theoretical literature so far. We will discuss three physical effects: the radiative decay, plasma oscillations, and retardation effects. We will show, by means of simple qualitative arguments and estimates, that these effects are evidently important under the conditions of experiments \cite{Mani02,Zudov03,Dorozhkin03,Yang03,Mani04,Studenikin04,Mani04a,Du04,Willett03,Mani03,Kovalev03,Studenikin04b}. We calculate the influence of microwaves on the electron distribution function 
and the microwave response of a finite-width 2D wire, accounting for electrodynamic effects, and show that these effects have a dominant role. We believe that this work may give another direction of thinking about the origin of microwave-induced phenomena in 2DESs.

\section{Radiative decay}\label{rd}

We begin the discussion with the effect of radiative decay. In this Section we will assume that the 2DES is uniform, occupies the plane $z=0$, infinite in $x$ and $y$ directions, and is placed in vacuum. Electromagnetic wave is assumed to be incident upon the 2DES along the $z$-axis. Physically, the radiative decay develops as the reaction of the medium (2DES) to radiation \cite{Jackson99}. Oscillating electric field of the incident wave forces 2D electrons to oscillate in the 2D plane, but oscillating 2D electrons emit a secondary radiation from the 2DES. As a result, the system loses energy, and the cyclotron-resonance line gets an additional contribution to the linewidth. The simplest way to calculate this contribution is to solve the Maxwell equations for electromagnetic waves passing through the 2DES. Such a solution \cite{Chiu76} gives for the transmission amplitude
\be 
t(\omega)=\frac 1{1+2\pi\sigma(\omega)/c}, 
\ee 
where $\sigma(\omega)$ is the conductivity of the 2DES and $c$ the velocity of light. If the 2DES is placed in a magnetic field and the electromagnetic wave is circularly polarized, we can substitute for $\sigma$ the Drude expression 
\be
\sigma(\omega)=
\frac{n_se^2}{m^\star} \frac i{\omega-\omega_c+i\gamma},
\ee
where $n_s$, $e$ and $m^\star$ are the density, the charge and the effective mass of 2D electrons, $\omega_c$ is the cyclotron frequency, and $\gamma$ is the scattering rate, related to the mobility $\mu=e/m^\star\gamma$. The transmission, reflection and absorption coefficients then get the form
\be
T=|t|^2=
1-\frac{\Gamma^2+2\gamma\Gamma}{(\omega-\omega_c)^2+(\gamma+\Gamma)^2},
\ee
\be
R=
\frac{\Gamma^2}{(\omega-\omega_c)^2+(\gamma+\Gamma)^2},
\ee
and
\be
A=
\frac{2\gamma\Gamma}{(\omega-\omega_c)^2+(\gamma+\Gamma)^2}.
\ee
One sees, that the linewidth of the cyclotron resonance here is determined by the collisional scattering rate $\gamma$ {\em plus} the electrodynamic contribution
\be
\Gamma=2\pi n_se^2/m^\star c.
\label{Gamma}
\ee
The ratio of the second contribution to the first one, 
\be
\frac \Gamma\gamma=\frac{2\pi n_se\mu}c=\frac{2\pi\sigma_0}c,
\ee
is much bigger than unity in the high-electron-mobility systems, used in the discussed experiments; here $\sigma_0=n_se\mu$ is the static conductivity of the 2DES. For a typical electron density of $n_s=3\times 10^{11}$ cm$^{-2}$ and for the mobility of $\mu\simeq 10^7$ cm$^2$/Vs, it is $\Gamma/\gamma\simeq 90$, and hence, the radiative decay effect is very important in the discussed phenomena. Notice, that at $\Gamma/\gamma\gg 1$ the incident-wave energy is mainly reflected by the 2D electron gas, but not absorbed in it.

Some electrodynamic properties of very clean 2D electron systems with $2\pi\sigma_0/c>1$ were studied in very interesting papers by Falko and Khmelnitskii \cite{Falko89}, and by Govorov and Chaplik \cite{Govorov89}. Related features of quantum-wire systems under the quantum-Hall conditions, where the similar relevant parameter has the form $2\pi/\rho_{xx}c$ and can be {\em much} larger than unity, were discussed in \cite{Mikhailov93} ($\rho_{xx}$ is the longitudinal resistivity of the 2DES). The effect of the electrodynamic line broadening is well known in the optics of metals and, for example, in the theory of powerful laser sources of very short electromagnetic bursts \cite{Kaplan02}. In 2DES samples, showing the microwave-induced zero-resistance effect, it has been recently directly observed in very important absorption experiments by Studenikin et al \cite{Studenikin04b}.

The formula (\ref{Gamma}) can be derived from simple physical considerations (some discussion of related effects can be also found in \cite{Mikhailov96a}). An incident electromagnetic wave forces 2D electrons to oscillate in phase with the frequency $\omega$. Each electron, oscillating relative to the positive background, produces a dipole radiation with the intensity \cite{Landau2} $I\sim \ddot d^2/c^3\sim \omega^4e^2a^2/c^3$, where $a$ is the oscillation amplitude and $d\sim ea$ is the dipole moment. The radiative decay rate $\Gamma_0$ of a {\em single} oscillating charge can then be determined dividing $I$ by its average energy $\sim m^\star\dot a^2\sim m^\star\omega^2 a^2$. This gives $\Gamma_0\sim e^2\omega^2/m^\star c^3$. For $N$ 2D electrons, oscillating in phase, the intensity $I$ should be multiplied by $N^2$, while the average energy -- by $N$, so one gets $\Gamma\sim N\Gamma_0$. The value of $N$ in the considered case is estimated as the number of electrons in the coherence area $\sim\lambda\times\lambda$, where $\lambda\sim c/\omega$ is the wavelength of radiation, so that $N\sim n_s\lambda^2$. This finally gives $\Gamma\sim N\Gamma_0\sim n_se^2/m^\star c$, in agreement with the exact formula (\ref{Gamma}). Equation (\ref{Gamma}) thus describes {\em coherent dipole} reradiation of electromagnetic waves by oscillating 2D electrons excited by microwaves. The quantity $\Gamma$ can be also treated as the probability of electron-{\em photon} scattering in the 2DES.

At $\Gamma\gg\gamma$ the radiative processes also govern the steady-state electron distribution function formed by microwaves. Incident electromagnetic radiation with the frequency close to the cyclotron one continuously excites electrons to higher Landau levels, and there must be a physical mechanism which returns the system back to the (quasi-)equilibrium. In some publications (e.g. \cite{Dmitriev03b}) inelastic-scattering processes due to electron-electron collisions are considered to be the reason of such relaxation. The probability of such inelastic processes $1/\tau_{in}$ is however much smaller than that of elastic processes ($\gamma$) \cite{Dmitriev03b}, while $\gamma$, in its turn, is much smaller than the probability of electron-photon scattering $\Gamma$. Under the experimental conditions ($\Gamma\gg\gamma\gg\tau_{in}^{-1}$) inelastic processes can thus be safely ignored, and the microwave-modified steady-state electron distribution function can be calculated as follows \cite{Mikhailov03c}.

Since the effect was observed under the quasiclassical conditions
\be
\hbar\gamma\ll kT\simeq \hbar \omega_c\lesssim \hbar\omega \ll E_F,
\label{conditions}
\ee
where $T$ is the temperature and $E_F$ is the Fermi energy, we can use the classical Boltzmann equation. If the sample is infinite and the external $ac$ field is uniform, it is written as
\be
\frac{\partial f}{\partial t}+\frac{\partial f}{\partial \bf p}\left(-e{\bf E}-\frac{e}c{\bf v}\times {\bf B}\right)=0,
\label{BE}
\ee
where we have ignored the scattering integral, since $\gamma\ll\omega$, $\omega_c$ and $\Gamma$. The electric field ${\bf E}$ here is {\em not} the electric field of the incident wave ${\bf E}_0$, but the total {\em self-consistent} $ac$ electric field at the plane $z=0$, related to the amplitude of the incident wave ${\bf E}_0$ by the Maxwell equations, $E=tE_0$. Eq. (\ref{BE}) has the exact solution 
\be
f({\bf p},t) = f_0({\bf p}-m^\star{\bf V}(t)),
\label{df}
\ee
valid at any amplitude of the electric field (i.e. in the nonlinear regime, too). Here $f_0$ is the Fermi-Dirac distribution function, ${\bf V}$ resolves the classical equations of motion for one electron, 
\be
V=\frac{eEe^{-i\omega t}}{im^\star(\omega-\omega_c)}=\frac{eE_0e^{-i\omega t}}{im^\star(\omega-\omega_c+i\Gamma)},
\label{sol}
\ee
and $V$ and $E$ are complex amplitudes of the velocity and field of the circularly polarized wave. One sees that radiative effects remove the divergency at $\omega=\omega_c$ even if the impurity and phonon scattering inside the sample is fully neglected, and that at $\Gamma\gg\gamma$ this is the most important relaxation mechanism.

\begin{figure}
\includegraphics[width=8.2cm]{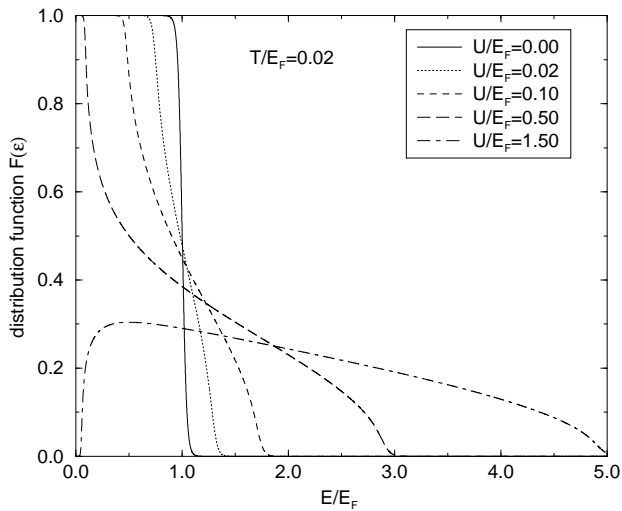}
\caption{Electron distribution function (\protect\ref{fdaveraged}) at different microwave power levels. The upper characteristic point on the energy axis (where $F$ approaches zero) corresponds to $E=(\sqrt{E_F}+\sqrt{U})^2$. The lower point (where $F$ approaches unity at $U<E_F$ and zero at $U>E_F$) corresponds to  $E=(\sqrt{E_F}-\sqrt{U})^2$.}
\label{fd}
\end{figure}

The function (\ref{df}) is periodic in time. The microwave-induced steady-state distribution function $F(\epsilon)$ is found by averaging $f({\bf p},t)$ over the oscillation period. This gives 
\be
F(\epsilon)\equiv\overline{f({\bf p},t)}=\frac 1\pi\int_0^\pi\frac {dx}{1+\exp\left[\frac{\epsilon -E_F+U+2\sqrt{U\epsilon}\cos x}{kT}\right]},
\label{fdaveraged}
\ee
where $U$ is proportional to the microwave power and resonantly depends on the frequency,
\be
U=\frac{e^2E_0^2}{2m^\star[(\omega-\omega_c)^2+\Gamma^2]}.
\ee
Figure \ref{fd} shows the function (\ref{fdaveraged}) at different values of $U/E_F$. At $U>E_F$ the function $F(\epsilon)$ describes inversion of population. This requires, however, rather strong microwave powers, which was probably not realized in the discussed experiments \cite{elecdyn-note1}.

\section{plasma oscillations and finite-size effects}

The second important effect which should be discussed in connection with the experiments \cite{Zudov01,Ye01,Mani02,Zudov03,Dorozhkin03,Yang03,Mani04,Studenikin04,Mani04a,Du04,Willett03,Mani03,Kovalev03,Studenikin04b} is plasma oscillations. Passing electromagnetic radiation through an {\em infinite} 2D electron gas, placed in a magnetic field, one would observe the cyclotron resonance at the frequency $\omega=\omega_c$. In a {\em finite-size} 2D sample with lateral dimensions $W$ the cyclotron resonance is shifted to a higher frequency due to a depolarization effect. The depolarization shift is always observed in far-infrared absorption spectra of quantum-dot and quantum-wire systems, see for instance \cite{Allen83,Demel88}. In quantum wires the resonance is seen at the bulk-magnetoplasmon frequency 
\be
\omega_{mp}= \sqrt{\omega_c^2+\omega_0^2},
\label{mp}
\ee 
where $\omega_0\approx\omega_p(\pi/W)$ is estimated as the 2D plasmon frequency
\be
\omega_p(q)=\sqrt{2\pi n_se^2q/m^\star\epsilon}
\label{omegap}
\ee
with the wavevector $q\sim\pi/W$, and $\epsilon$ is the dielectric constant of surrounding medium. The same result also follows from the generalized Kohn theorem \cite{Shikin92}. 

In macroscopic samples at microwave frequencies one should also expect the influence of depolarization (plasma) effects. As seen from (\ref{mp})--(\ref{omegap}), the plasma shift can be neglected, if the sample size or the frequency are sufficiently large, $\omega\gg\omega_0$, or $\omega^2W\gg 2\pi^2n_se^2/m^\star\epsilon$. Table \ref{tab1} shows that this condition was {\em not} satisfied in the discussed experiments: in almost all the cases the microwave frequency was comparable with or smaller (sometimes much smaller) than the plasma frequency, $\omega\lesssim\omega_0$. The finite-size and plasma effects should thus be included in the theory. It should be emphasized that, in the very similar microwave photoresistance experiment \cite{Vasiliadou93} made with a one-order-of-magnitude lower mobility sample ($\mu\simeq 10^6$ cm$^2$/Vs) the magnetoplasmon resonance at the frequency (\ref{mp}) was really observed. However, the question, why the depolarization shift was seen in the old, lower-mobility samples, but is not seen in the new, very-high-mobility samples, or vice versa, why the new effect of giant photoresistance oscillations was not seen in \cite{Vasiliadou93}, has not been answered yet.

\begin{table}
\caption{\label{tab1} Comparison of the microwave frequency $f=\omega/2\pi$ and the plasma frequency $f_0=\omega_0/2\pi$ in several zero-resistance-states experiments.}
\begin{ruledtabular}
\begin{tabular}{ccccc}
Ref. & $n_s/$cm$^2$ 
& size & $f_0$ (GHz) 
& $f$ (GHz) \\
\hline
\cite{Mani02} & $3\times 10^{11}$ & $50-200$ $\mu$m  & $64-128$  & $27-115$ \\
\cite{Zudov03} & $3.5\times 10^{11}$  & $\sim 5$ mm & $\sim 14$  & $30-150$ \\
\cite{Dorozhkin03} & $3\times 10^{11}$ & 200 $\mu$m & 64  & $10-170$  \\
\cite{Willett03} & $2\times 10^{11}$  & $0.4-5$ mm  & $10-37$   & $7-20$  \\
\end{tabular}
\end{ruledtabular}
\end{table}

Physically, the depolarization shift arises due to the difference between the total self-consistent $ac$ electric field $E$, really acting on the 2D electrons at the plane $z=0$, and the external field of the incident electromagnetic wave $E_0$. As was seen in the previous Section, in an infinite sample this difference leads to a substantial broadening of the cyclotron-resonance line. In a finite-size sample it is even more important as it shifts the resonance frequency itself. For a 2D wire of the width $W$ the relation between $E$ and $E_0$ has the form (see e.g. \cite{Mikhailov98c})
\be E=\frac{E_0}{\zeta(\omega)}, \ \ {\rm where} \ \  \zeta(\omega)=1-\frac{\omega_0^2}{\omega^2-\omega_c^2} \label{fieldrelation}\ee
is a screening function (we have ignored the linewidth here).

In so far published theoretical scenarios only infinite-size 2D systems have been considered ($W\to\infty$, $\omega_0\ll \omega$), and photoresistance oscillations associated with the cyclotron frequency and its harmonics have been obtained. If to assume that the same models work in a real, finite-size sample, than one should evidently expect, in view of Eq. (\ref{fieldrelation}), that these oscillations are shifted to magnetoplasmon frequencies. As seen from the above estimates and Table \ref{tab1}, this depolarization shift is very far from being negligible, and hence a proper explanation of the discussed phenomena is still absent.

Another (indirect) evidence of the importance of plasma oscillations in microwave-induced magnetotransport phenomena can be found in a recently published paper \cite{Kukushkin-unpubl}. In this paper, another type of photoresistance oscillations has been observed. Although these oscillations, contrary to \cite{Mani02,Zudov03,Dorozhkin03,Yang03,Mani04,Studenikin04,Mani04a,Du04,Willett03,Mani03,Kovalev03,Studenikin04b}, are periodic in $B$ and have been found in the opposite magnetic-field regime $\omega\lesssim\omega_c$, they are quite similar in appearance to the $1/B$-periodic oscillations observed at $\omega\gtrsim\omega_c$ in \cite{Mani02,Zudov03,Dorozhkin03,Yang03,Mani04,Studenikin04,Mani04a,Du04,Willett03,Mani03,Kovalev03,Studenikin04b}: the photoresistance oscillate around the dark-resistance curves, and at sufficiently large microwave powers approaches zero. The $B$-periodic oscillations \cite{Kukushkin-unpubl}, however, were shown to be due to the excitation of plasma waves in the sample (the edge magnetoplasmons, relevant in the regime $\omega\lesssim\omega_c$). Accordingly, it seems to be reasonable to expect that the bulk magnetoplasmons (\ref{mp}) (or maybe magnetoplasmon-polaritons, see the next Section) play a part in the phenomena observed in \cite{Zudov01,Ye01,Mani02,Zudov03,Dorozhkin03,Yang03,Mani04,Studenikin04,Mani04a,Du04,Willett03,Mani03,Kovalev03,Studenikin04b}.

\section{retardation effects}

\begin{figure}[tph!]
\includegraphics[width=8.2cm]{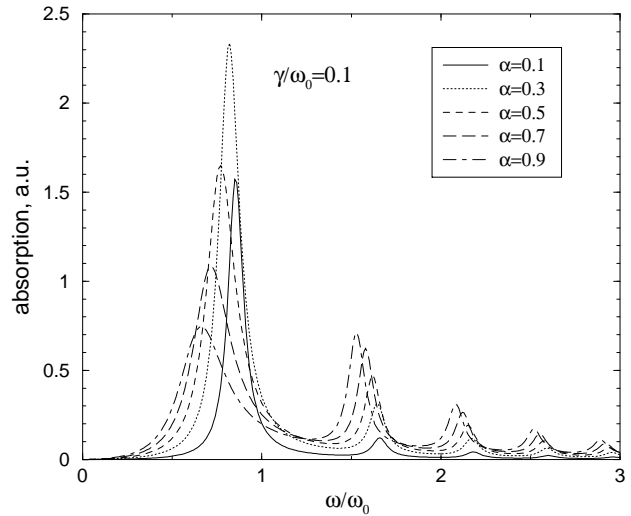}
\caption{Absorption spectrum of a 2D wire with the width $W$ at different values of the retardation parameter $\alpha$.}
\label{abs}
\end{figure}

The third electrodynamic effect important in the discussed experiments 
is retardation. The retardation effects become essential when the velocity of 2D plasmons (\ref{omegap}) approaches the velocity of light $c/\sqrt{\epsilon}$. The relevant dimensionless parameter has the form 
\be
\alpha=\sqrt{\frac{2n_se^2W}{m^\star c^2}}\approx 0.29\sqrt{n_s[10^{11}\ {\rm cm}^{-2}]\times W[{\rm mm}]},
\label{alpha}
\ee
where we have used $q\sim\pi/W$. In microscopic electron systems like quantum wires and dots, this parameter is small as compared to unity, and the resonant response frequencies can be calculated in the quasistatic approximation, using the Poisson equation. In macroscopic 2DES samples with lateral dimensions of order of 1 mm, $\alpha$ is comparable with or larger than unity, and retardation effects should be included in the theory (for instance, $\alpha\sim 1.21$ in \cite{Zudov03}). It was recently shown in \cite{Kukushkin03a,Kukushkin03b}, that retardation effects essentially modify the absorption spectra of 2DES at $\alpha\simeq 1$, especially at $\omega\gtrsim\omega_c$. To give an idea of how the retardation may influence electromagnetic response of a finite-size 2DES (full report on this study will be published elsewhere), we show in Figure \ref{abs} the absorption spectrum of a 2D wire with the width $W$ ($z=0$, $|x|<W/2$), at different values of the $\alpha$-parameter. 
One sees that, at small $\alpha$ (the quasistatic limit) the absorption spectrum is similar to that of quantum wires: a strong peak corresponding to the fundamental 2D plasmon mode with $q\sim\pi/W$ is accompanied by a number of very weak higher-harmonics peaks, corresponding to $q\sim (2n+1)\pi/W$ with $n=1,2,\dots$ (the Kohn theorem is not applicable here as the confining potential is not purely parabolic). Increasing $\alpha$ leads to essential modifications of the absorption spectrum. The fundamental mode first increases, and then decreases in amplitude, additionally experiencing a red-shift. The amplitude of this mode is maximum if the radiative losses are equal to the dissipative ones. The behaviour of higher modes is different. Their frequency continuosly decreases and their amplitude continuously increases with the growth of $\alpha$. At $\alpha\simeq 1$ the absorption spectrum exhibits many modes with almost equal amplitude, in contrast to the quasistatic limit. When the magnetic field is on, all these modes increase in frequency, which leads to a many-peak dependence of the absorption spectra on $B$. Such behaviour was observed in \cite{Kukushkin03a,Kukushkin03b} and was qualitatively described by Stidenikin et al at the end of Section III B in \cite{Studenikin04}. 

As seen from Figure \ref{abs}, retardation effects uncover the higher 2D plasmon harmonics hardly observable in the quasistatic limit. We expect that they similarly influence the higher cyclotron resonance harmonics (the Berstein modes) at $\omega=m\omega_c$, $m=2,3\dots$, which are usually weak in the quasistatic regime. This may explain a weak dependence of microwave induced photoresistance oscillations on the cyclotron-harmonics index $m$.

\section{conclusions}

We have argued that electrodynamic effects, not considered in theoretical literature so far, may be crucial for understanding the physics of microwave induced magnetotransport phenomena recently observed in 2DESs. Results of more detailed study of these effects will be reported.


I am grateful to Ramesh Mani, Sergey Dorozhkin, Igor Ku\-kush\-kin, Jurgen Smet and Klaus von Klitzing for numerous discussions of experimental details and related physical issues, as well as Hans-Erik Nilsson and Sverker Edvardsson for interest to this work.


\begin{thebibliography}{10}
\bibitem{elecdyn-note1}
The microwave field in the discussed experiments was of order of $1-5$ V/cm.
  Notice that in \protect\cite{Lei03a} it was assumed to be almost two orders
  of magnitude larger.
\end{thebibliography}

\end{document}